\documentclass[twocolumn,trackchanges]{aastex701}
\usepackage{amsmath}
\newcommand{\rev}[1]{\textbf{#1}}

\newcommand {\hMpc}    {\ $h^{-1}\,\mathrm{cMpc}$}
\newcommand {\hkpc}    {\ $h^{-1}\,\mathrm{ckpc}$}
\newcommand {\hM}   {\ $h^{-1} \  M_{\odot}$}

\begin{document}

\title{The Origin of Spin-Alignment of Dark Matter Subhalos}

\author[orcid=0009-0000-2097-5567]{Daiki Osafune}
\affiliation{Department of Cosmosciences, Graduate School of Science, Hokkaido University, N10 W8, Kitaku, Sapporo, 060-0810, Japan}
\email[show]{osafune@astro1.sci.hokudai.ac.jp}  

\author[orcid=0000-0002-8779-8486]{Keiichi Wada}
\affiliation{Graduate School of Science and Engineering, Kagoshima University, Kagoshima 890-0065, Japan}
\affiliation{Research Center for Space and Cosmic Evolution, Ehime University, Matsuyama 790-8577, Japan}
\affiliation{Department of Physics, Faculty of Science, Hokkaido University, N10 W8, Kitaku, Sapporo 060-0810, Japan}
\email{wada@astrophysics.jp}

\author[orcid=0000-0002-5316-9171, gname='Tomoaki', sname='Ishiyama']{Tomoaki Ishiyama}
\affiliation{Digital transformation enhancement council, Chiba University, 1-33, Yayoi-cho, Inage-ku, Chiba, 263-8522, Japan}
\email{ishiyama@chiba-u.jp}

\author[orcid=0000-0003-0137-2490, gname='Takshi', sname='Okamoto']{Takashi Okamoto}
\affiliation{Department of Physics, Faculty of Science, Hokkaido University, N10 W8, Kitaku, Sapporo 060-0810, Japan}
\email{takashi.okamoto@sci.hokudai.ac.jp}  


\begin{abstract}
Subhalo spin is essential for modeling galaxy formation and controlling systematic uncertainties in intrinsic alignment (IA) studies. However, the physical mechanisms governing subhalo spin acquisition within the tidal environments of host halos remain poorly understood. In this work, we investigate the alignment between subhalo and host halo spins using the high-resolution cosmological $N$-body simulation, Shin-Uchuu.
We find that the spin alignment between subhalos and host halos becomes increasingly pronounced toward the central regions. Our analysis reveals that subhalos typically acquire spin in the same direction as their orbital angular momentum. Since the orbital angular momentum of most subhalos is aligned with the host halo spin, an overall alignment between subhalo and host spins emerges. When classified by orbital orientation, however, subhalo spins in the inner regions are found to be oriented perpendicularly or anti-parallel to the host spin for polar and retrograde orbits, respectively. These results provide strong evidence that subhalo spins are acquired through torques exerted by the tidal field of the host halo. Furthermore, we demonstrate that the mass ratio and the radial distance from the host center are the primary parameters governing subhalo spin alignment, while the dependence on the accretion redshift plays a less significant role compared to the radial distance and mass ratio.
\end{abstract}



\section{Introduction}
In the standard cosmological model, halos form via gravitational instability of initial dark matter density fluctuations and grow through the merger and accretion of other halos.
When a smaller halo accretes onto a more massive halo, it is referred to as a subhalo, while the larger system is termed its host halo.
Subhalos are subsequently tidally stripped and disrupted by the tidal field of the host halo \citep{2005MNRAS.359.1029V,2008MNRAS.383...93B}.
The presence of subhalos is observable through gravitational lensing \citep{2009ApJ...693..970N} and the kinematics of satellite galaxies \citep{2012ApJ...748...20C}.

Weak gravitational lensing serves as a powerful tool for probing the mass distribution and internal structures of galaxies \citep{
2014ApJ...784...90O, 2015MNRAS.448.2704I, 2018MNRAS.478.1244S, 2018ApJ...862....4L,2019PASJ...71...43H,2024A&A...689A.298G,2025arXiv251209342F}.
However, a significant systematic challenge in these studies is the intrinsic alignment (IA), which is the intrinsic correlation of galaxy shapes and orientations with the surrounding gravitational field.
Understanding IA is crucial for ensuring the precision and reliability of weak lensing surveys \citep[e.g.,][for a review]{2015SSRv..193....1J}.
Especially on small scales ($\lesssim 1{\rm Mpc}$), accounting for the IA of subhalos and satellite galaxies within central galaxy halos improves the precision of the analysis \citep{2024OJAp....7E..45V, 2023ApJ...957...45X}.
For early-type galaxies, the intrinsic alignment of galaxies is often described by models based on the linear alignment model \citep{2004PhRvD..70f3526H, 2022MNRAS.514.2049Z}. This model characterizes the correlation between the observed ellipticity of a galaxy and the local gravitational tidal shear field.
In contrast, for late type galaxies and subhalos, models that account for the influence of tidal torques induced by the surrounding tidal field have been proposed \citep{2008ApJ...681..798L,2015MNRAS.452.3369C, 2025ApJ...995L...8M, 2025A&A...703A..67M}.
The intrinsic alignment of galaxy shapes depends on the surrounding environment, ranging from the large-scale structure to the internal distribution within the host halo \citep{2024A&A...688A..40R,2025A&A...699A.215R} 
In particular, for subhalos within host halos, it is thought that tidal forces influence the orientation of the major axis, and observational and theoretical studies discovered a tendency for the major axis of subhalos to point towards the centre of the host halo \citep{2008ApJ...672..825P,2008MNRAS.388L..34K,2020MNRAS.495.3002K,2019MNRAS.484.4325W}. 
 Although this trend is generally attributed to tidal torques, its detailed physical mechanism is not clear.

Previous research suggests that the effect of tidal torques from the host halo extends not only to the shape of subhalos but also to their spin. Theoretical studies have shown a tendency for the spin vectors of subhalos to align with those of their host halos \citep{2004MNRAS.352..376A,2018A&A...613A...4W, 2025MNRAS.538..963M} and these results further suggest that this spin alignment is related to the properties of satellite galaxies, such as their color, morphology, and stellar mass, implying a connection with their formation histories. 
Consistently, similar spin alignments have also been identified in observations of galaxy groups \citep{2025ApJ...992L..17W}, further implying a fundamental connection between tidal interactions and galaxy formation histories.
Despite these findings, the specific physical origin of subhalo spin alignment remains elusive, largely due to the insufficient mass resolution and limited sample sizes of existing studies. For example, \cite{2018A&A...613A...4W} analyzed a stacked sample within $0.3 < z < 0.8$ with a box size of $L=100$\hMpc, which may not provide the detailed small-scale dynamics or statistical robustness required for a comprehensive investigation.

In this study, we investigate the statistical properties induced by tidal torques from host halos using snapshots of the high-resolution cosmological $N$-body simulation, Shin-Uchuu which features a box size of $L=140$\hMpc\ and 100 times higher dark matter mass resolution than the simulation used in \cite{2018A&A...613A...4W}.
The improvements in both resolution and sample size now allow for robust statistical investigations of subhalos, even after imposing multiple parameter constraints. By analyzing the relationship between subhalo spin and the distance from the host halo center under constraints of some parameter, this study aims to elucidate how the tidal field of the host halo influences subhalo properties.

This paper is organised as follows. Details of our simulations, sample selection, and analysis method are described in Section 2. In Section 3, we present the statistical results on the alignment of subhalo spins and orbital angular momenta. Finally, we provide a discussion and conclusions in Section 4.

\section{Data and Analysis}
\subsection{Cosmological N-body simulation data}

We use a cosmological $N$-body simulation, Shin-Uchuu \citep{2021MNRAS.506.4210I}. The cosmological simulation consists of $N=6400^3$ dark matter particles in a comoving cubic box of size $L=140.0$\hMpc. 
The particle mass is $m_{\rm p}=8.97\times10^{5}$\hM\ and the softening length is \textbf{$\epsilon = 0.4$\hkpc}.
The initial conditions were generated using the 2LPTIC code  \citep{2006MNRAS.373..369C}  \footnote{\url{https://cosmo.nyu.edu/roman/2LPT/}}, which adopts second-order Lagrangian perturbation theory. The simulation was evolved from an initial redshift of $z=127$ to $z=0$ using the massively parallel TreePM code, GREEM\footnote{\url{https://hpc.imit.chiba-u.jp/~ishiymtm/greem/}} \citep{2012arXiv1211.4406I,2009PASJ...61.1319I}.
The cosmological parameters are $\Omega_{0} = 0.3089, \Omega_{b} = 0.0486, \lambda_{0} = 0.6911, h = 0.6774, n_{s} = 0.9667$, and $\sigma_{8} = 0.8159$ \citep{2020A&A...641A...6P}.

Halos/subhalos are identified by ROCKSTAR\footnote{\url{https://bitbucket.org/gfcstanford/rockstar/}} halo finder \citep{2013ApJ...762..109B}, which outputs physical properties such as position, velocity and angular momentum of halos. 
The angular momentum of the subhalo (or host halo) $\vec{J}_{\rm sub(host)} $ is calculated using all particles determined to be gravitationally bound by the halo finder.
Let $m_{i},\vec{r}_i$ and $\vec{v}_i$ denote the mass, position, and velocity of the $i$-th bound particle relative to the center of the halo, respectively. The angular momentum is then defined as:

\begin{equation}
	\vec{J}_{\rm sub(host)} 
	=
	\sum^{N}_{i=1} m_i
		\left(
			\vec{r}_i\times \vec{v}_i
		\right)
\end{equation}

\begin{figure*}[ht!]
	\centering
	\includegraphics[width=2\columnwidth]{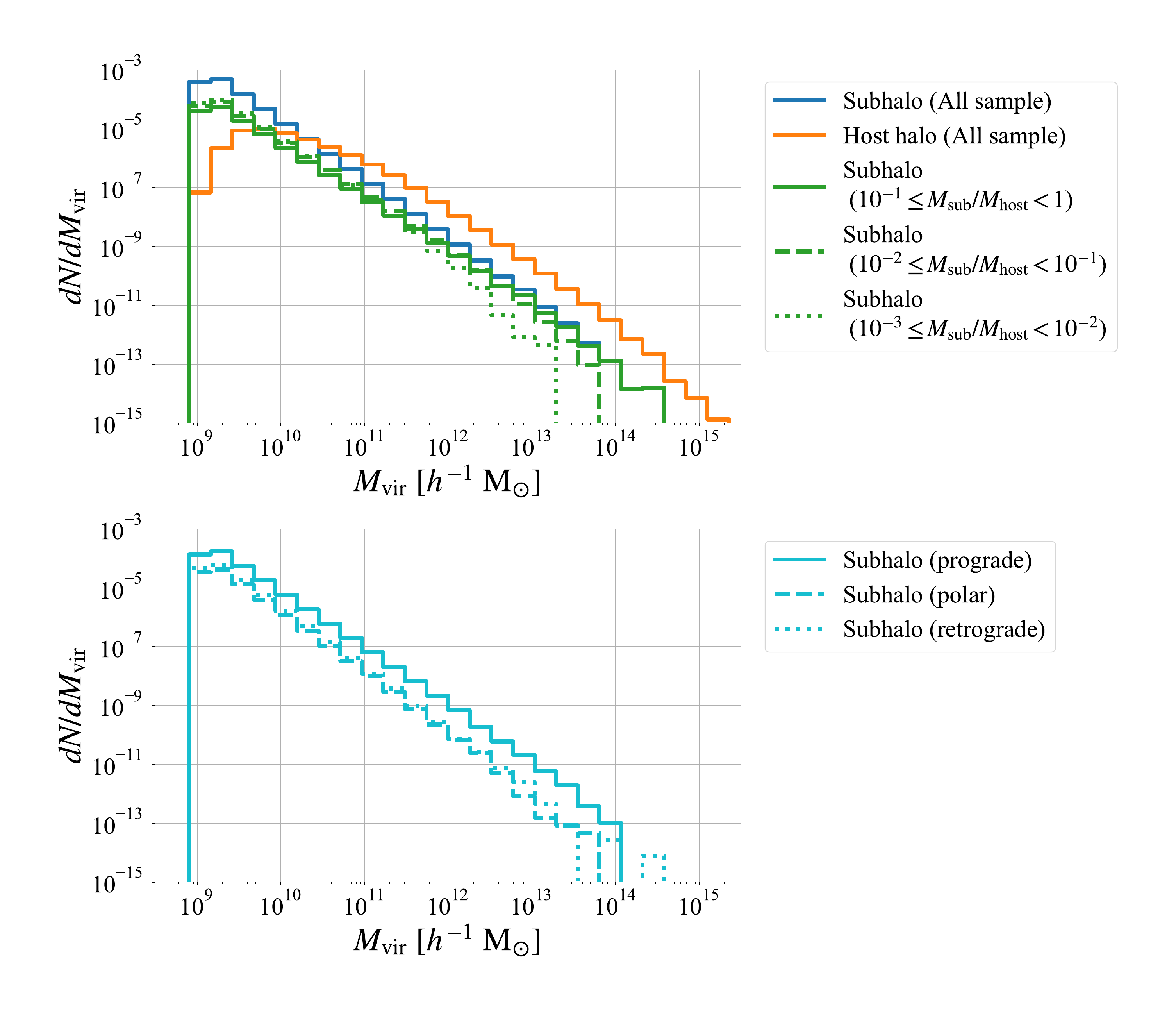}
	\caption{Sample sizes and mass distributions of subhalos and their host halos. The panels show differential mass histograms for $M_{\rm sub}$ and $M_{\rm host}$. \rev{In the top panel}, ``host halos'' are defined as those hosting the subhalos in each sample and are counted without duplication.
\rev{Top} : Full sample of subhalos (blue) and their host halos (orange), along with subsamples of subhalos categorized by the subhalo-to-host mass ratio: $10^{-1} \leq M_{\rm sub}/M_{\rm host} < 1$, $10^{-2} \leq M_{\rm sub}/M_{\rm host} < 10^{-1}$, and $10^{-3} \leq M_{\rm sub}/M_{\rm host} < 10^{-2}$ (green).
\rev{Bottom} : Subsamples categorized by the orbital orientation of subhalos (prograde, polar, and retrograde), show with solid, dashed and dotted lines (cyan). The definition of orbital orientation is given in \S\ref{subsec:orbital}. 
} 
	\label{fig:HistSample}
\end{figure*}

The merger trees are constructed by CONSISTENT TREES\footnote{\url{https://bitbucket.org/pbehroozi/consistent-trees/}} \citep{2013ApJ...763...18B} based on the data that output from ROCKSTAR at each redshift. The halo and subhalo catalogues are available on the Skies \& Universes site.\footnote{\url{https://www.skiesanduniverses.org/Simulations/Uchuu/}} 
We analyze subhalos at $z=0$ that consist of at least 1,000 dark matter particles and reside in host halos that are not themselves subhalos of any other halo. In other words, our analysis excludes "sub-subhalos" or any further levels of the subhalo hierarchy.
This selection ensures the numerical robustness of measured physical properties and avoids the influence of complex dynamics within nested hierarchical structures.
The sample of subhalos at $z=0$ used in this study consists of $N_{\rm sub} = 1.13\times 10^6$, with masses ranging from $M_{\rm sub,min}=8.97\times 10^8$\hM\ to $M_{\rm sub,max}=1.69\times 10^{14}$\hM. Similarly, the host halo sample includes $N_{\rm host}=3.07\times 10^5$, with masses ranging from $M_{\rm host,min}=9.26\times 10^{8}$\hM to $M_{\rm host,max}=1.23\times 10^{15}$\hM. 
Figure \ref{fig:HistSample} shows the differential for the subhalo samples and their respective host halos, along with subsamples categorized by mass ratio \rev{$M_{\rm sub}$ and $M_{\rm host}$} and orbital orientation.

The subsamples categorized by the mass ratio $M_{\mathrm{sub}}/M_{\mathrm{host}}$ contain comparable numbers of objects in each bin, although differences in sample size become apparent for $M_{\mathrm{vir}} > 10^{12}\,h^{-1}M_\odot$.
For the orbital-orientation subsamples, the number of subhalos in prograde orbits is larger by a factor of 3--4 than that in retrograde or polar orbits, while the retrograde and polar subsamples contain comparable numbers of objects.

\begin{figure}[ht!]
	\includegraphics[width=\columnwidth]{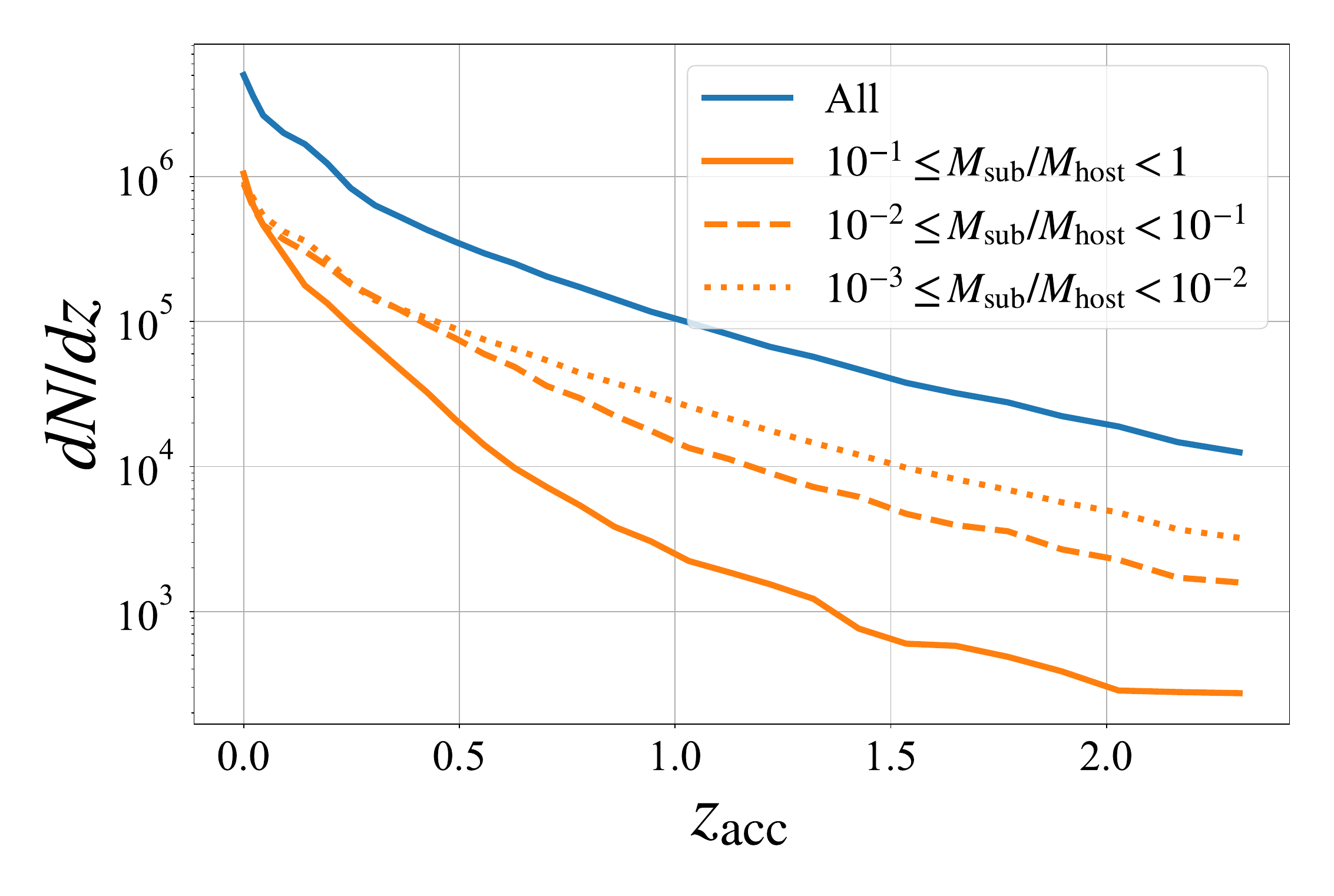}
	\caption{Differential histogram of subhalos as a function of the accretion redshift, $z_{\rm acc}$. The blue line shows the full sample, while the orange lines represent subsamples categorized by the subhalo-to-host mass ratio, $M_{\rm sub}/M_{\rm host}$. Subhalos are included only if the descendant of their host halo at $z_{\rm acc}$ corresponds to the host halo at $z=0$. The number of subhalos decreases toward higher $z_{\rm acc}$, particularly for higher mass-ratio subsamples.}
	\label{fig:HistSampleZacc}
\end{figure}

We investigate the dependence on accretion timing by analyzing subhalos as a function of the accretion redshift, $z_{\rm acc}$.
The accretion redshift is determined from the subhalo’s main branch of the merger tree, which traces the most massive progenitor at each timestep.
We define $z_{\rm acc}$ as the earliest snapshot at which the subhalo is identified as a member of a host halo, provided that it remains continuously associated with that host until $z=0$.
Subhalos are excluded if the descendant of their host halo at $z_{\rm acc}$ does not correspond to the host halo at $z=0$.
Figure \ref{fig:HistSampleZacc} shows the differential histogram of subhalos as a function of $z_{\rm acc}$. The sample is divided into three mass-ratio bins: $10^{-3} \le M_{\rm sub}/M_{\rm host} < 10^{-2}$, $10^{-2} \le M_{\rm sub}/M_{\rm host} < 10^{-1}$, and $10^{-1} \le M_{\rm sub}/M_{\rm host} < 1$.
The sample size in the range $0 \le z_{\rm acc} < 0.5$ is approximately $100{,}000$ in each mass-ratio bin. At higher accretion redshifts, the number of subhalos decreases with increasing mass ratio, ranging from approximately $20{,}000$ to $3{,}000$ for $0.5 < z_{\rm acc} < 1$ and from $10{,}000$ to $1{,}000$ for $1 < z_{\rm acc} < 2$.

\begin{figure*}[ht!]
	\centering
	\includegraphics[width=2\columnwidth]{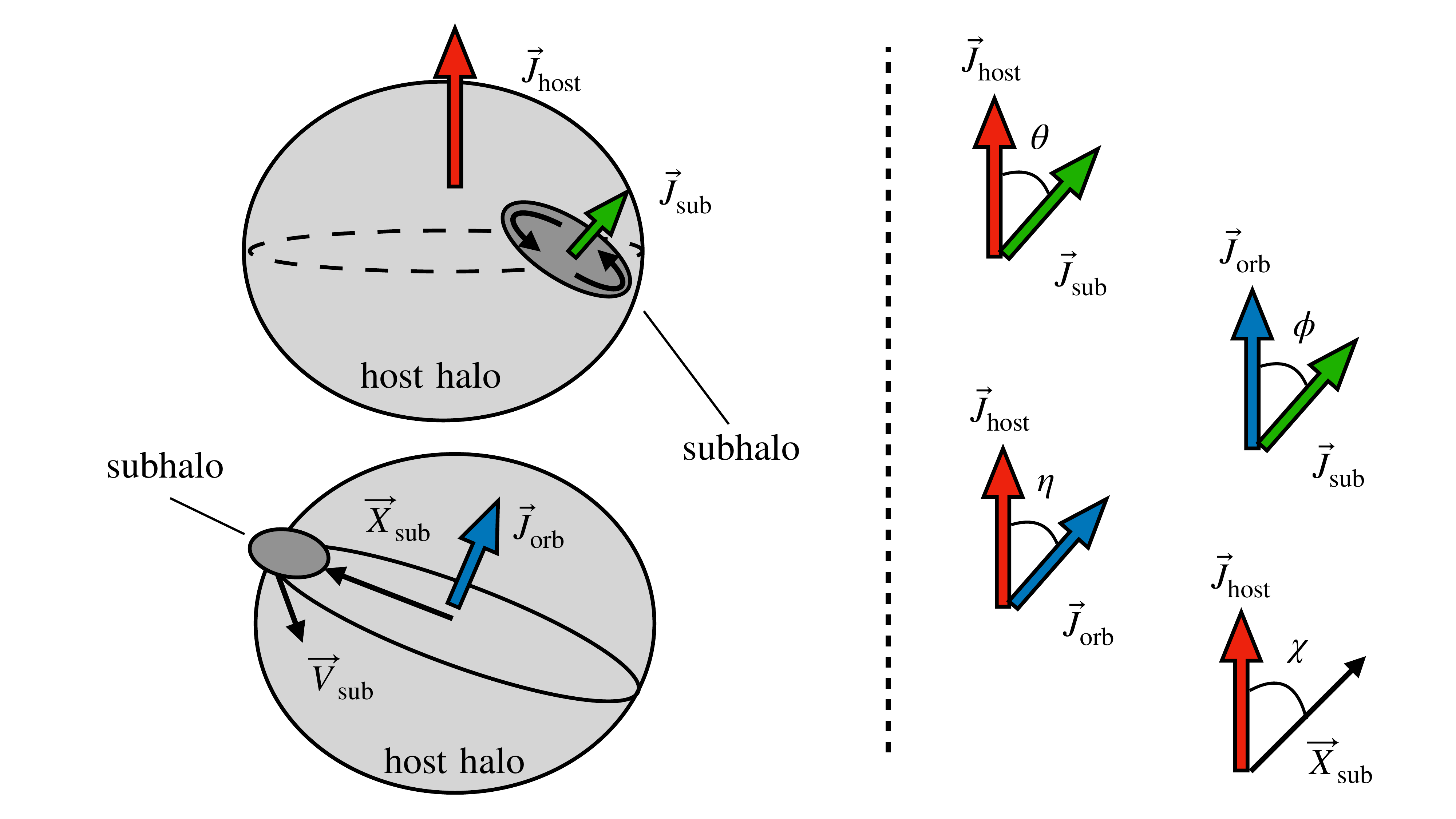}
	\caption{Schematic illustration of the vectors and the alignments of subhalo and host halo used for this study.} 
	\label{fig:image}
\end{figure*}

To investigate the various alignments, we will analyze the cosine values of the angles defined by the following quantities:
\begin{itemize}
	\item The alignment between the spin of subhalos and their host halos ($\cos\theta$).
	\item The alignments between the orbital angular momentum of the subhalos, $\vec{J}_{\rm orb}$ and the spin of both the subhalos ($\cos\phi$) and the host halos ($\cos\eta$).
	\item The positional alignment of subhalos relative to the spin direction of host halo ($\cos\chi$).
\end{itemize}
These alignments are illustrated in Figure \ref{fig:image} and defined by the following equations:

\begin{align}
	\cos\theta
	\equiv 
	\frac{{\vec{J}_{\rm sub}}\cdot {\vec{J}_{\rm host}}}
		{\left|{\vec{J}_{\rm sub}}\right|\left|{\vec{J}_{\rm host}}\right|},\\
	\cos\phi
	\equiv
	\frac{{\vec{J}_{\rm sub}}\cdot {\vec{J}_{\rm orb}}}
		{\left|{\vec{J}_{\rm sub}}\right|\left|{\vec{J}_{\rm orb}}\right|},\\
	\cos\eta
	\equiv
	\frac{{\vec{J}_{\rm orb}}\cdot {\vec{J}_{\rm host}}}
		{\left|{\vec{J}_{\rm orb}}\right|\left|{\vec{J}_{\rm host}}\right|},\\
	\cos\chi
	\equiv
	\frac{{\vec{X}_{\rm sub}}\cdot {\vec{J}_{\rm host}}}
		{\left|{\vec{X}_{\rm sub}}\right|\left|{\vec{J}_{\rm host}}\right|},
\end{align}
where $(\cdot)$ denotes the inner product operation.
The orbital angular momentum of subhalos, $\vec{J}_{\rm orb}$ is calculated from the subhalo's position ($\vec{X}_{\rm sub}$), mass ($M_{\rm sub}$), and velocity ($\vec{V}_{\rm sub}$) relative to the center of the host halo, which are output by ROCKSTAR:
\begin{equation}
	\vec{J}_{\rm orb}
	=
	M_{\rm sub}(\vec{X}_{\rm sub} \times \vec{V}_{\rm sub})
\end{equation}
Using these quantities, and the distance from the center of the host halo normalised by its virial radius ($R_{\rm vir}$), we investigate the influence of the host halo on the alignment of the subhalos.

To evaluate these alignment trends, we introduce the excess probability density, $1+\xi$.
This quantity represents the probability distribution relative to a uniform distribution, defined as:
\begin{equation}
1 + \xi = \frac{P\left({\rm Data}\right)}{P_{\rm uni}}
\end{equation}
where $P\left({\rm Data}\right)$ and $P_{\rm uni}$ are the measured and the uniform probability distributions of the alignments, respectively.
In this representation, a value of $1+\xi > 1$ indicates a statistical preference for alignment, while $1+\xi = 1$ at all values of the variable corresponds to a uniform distribution.

\section{Results}
\subsection{Angle between angular momentum of subhalo and host halo}

\begin{figure*}[ht!]
	\centering
	\includegraphics[width=2\columnwidth]{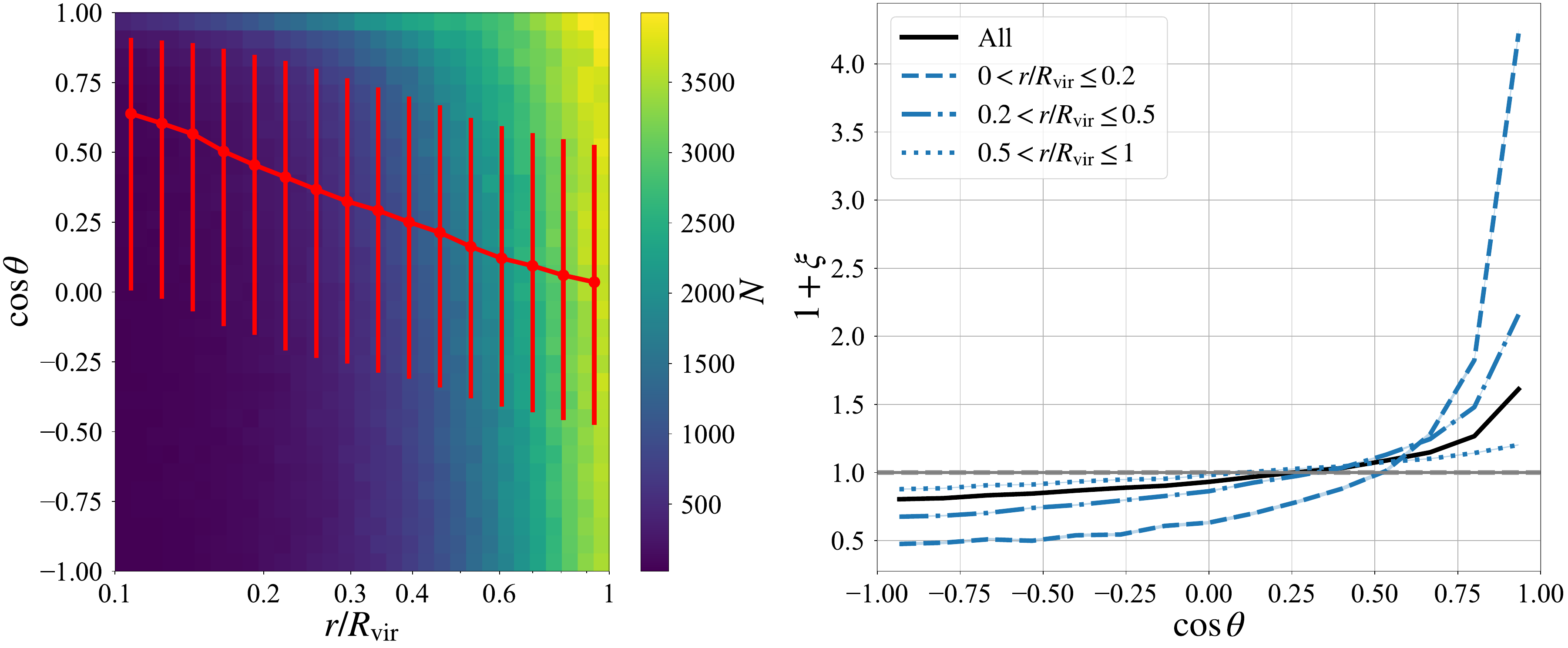}
	\caption{Left : A 2D histogram of the angle between the spin of the subhalos and the host halos, $\cos\theta$, and the distance from the center of the host halos, $r/R_{\rm vir}$. The red lines indicate the median, and the 25th and 75th percentiles. Right : excess probability density of $\cos\theta$, y-axis shows the deviation from a uniform probability distribution. Data are the full sample (black line) and subsample by $r/R_{\rm vir}$ (blue lines). The shaded regions represent Poisson errors, which are negligibly narrow due to the large sample size.}
	\label{fig:SAbasic}
\end{figure*}

Figure \ref{fig:SAbasic} shows the distribution of the distance from the center of the host halos, $r/R_{\rm vir}$, and the spin alignment between subhalo and host halo, $\cos\theta$. 
The left panel presents a two-dimensional (2D) histogram of these quantities, with the median and the interquartile range (25th to 75th percentiles) of $\cos\theta$ overlaid for each radial bin to highlight the average trend.
At the outskirts of the host halo, the median value of $\cos \theta$ is close to zero, and interquartile range (25th to 75th percentiles) is approximately $\pm 0.5$, indicating little to no preferred alignment. However, both the median and the overall distribution of $\cos \theta$ increase toward the center, demonstrating a clear tendency for subhalo spin directions to become more aligned with that of the host halo in the inner regions.

The right panel of Figure \ref{fig:SAbasic} shows the excess probability density function of $\cos\theta$, where the values on the vertical axis represent the excess probability relative to a uniform distribution. We present results for the full sample and three subsamples binned by their distance from the host halo center $r/R_{\rm vir}$. The full sample reveals a statistical tendency for subhalo and host halo spins to align. Furthermore, the alignment is close to a uniform distribution at the outskirts of host halos, whereas it becomes progressively stronger toward the central regions.
This is consistent with the results of previous studies \citep{2004MNRAS.352..376A, 2018A&A...613A...4W}, which focus on the spin alignment between the satellite and host galaxies.

\begin{figure*}[ht!]
	\centering
	\includegraphics[width=2\columnwidth]{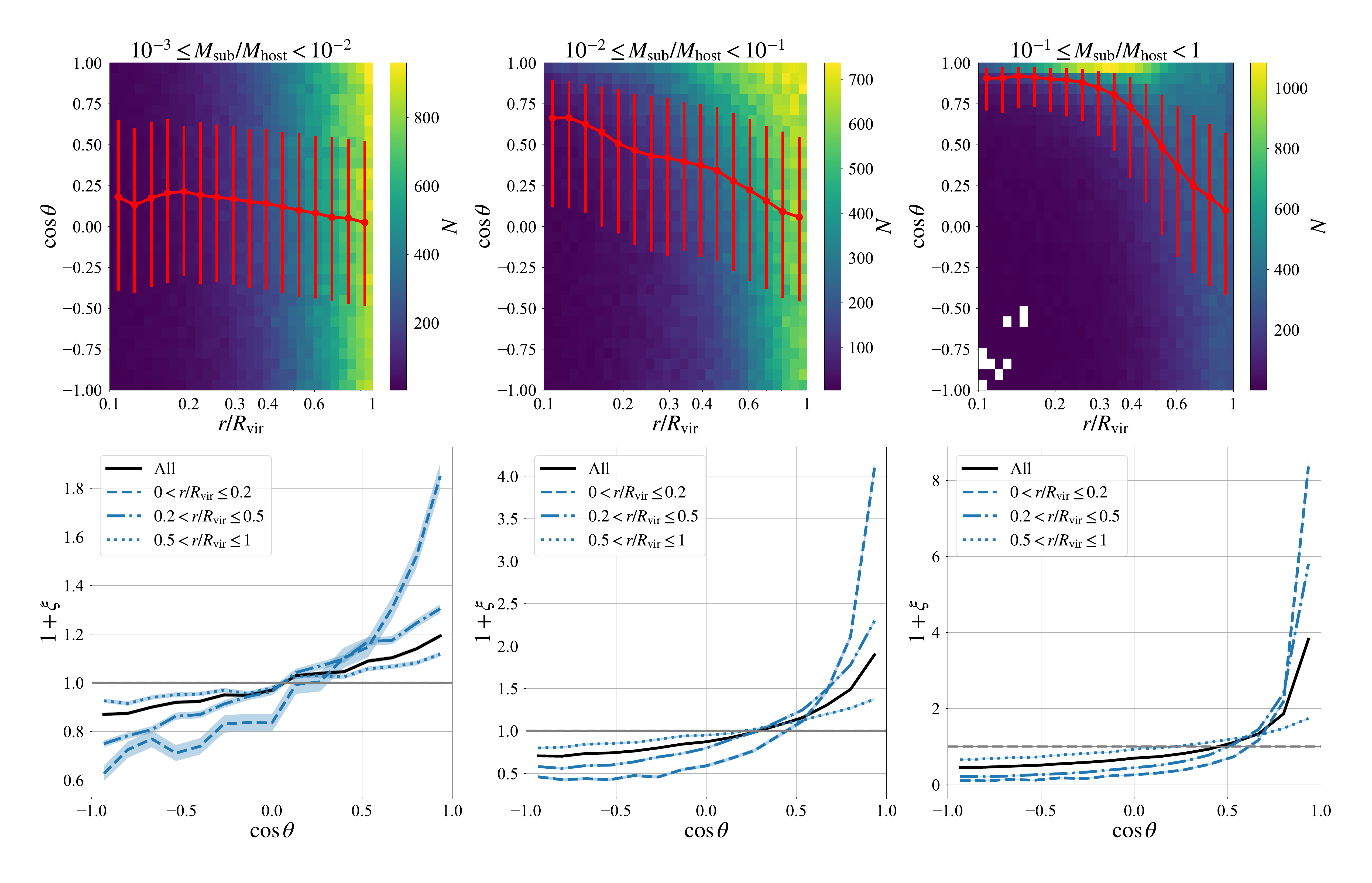}
	\caption{Upper panels: 2D histograms of $\cos\theta$ and $r/R_{\rm vir}$. Lower panels: excess probability density of $\cos\theta$.
	Each column represents a subsample restricted by mass ratio: $10^{-3} \le M_{\rm sub}/M_{\rm host} < 10^{-2}$ (left), $10^{-2} \le M_{\rm sub}/M_{\rm host} < 10^{-1}$ (middle), and $10^{-1} \le M_{\rm sub}/M_{\rm host} < 1$ (right). Shaded regions represent Poisson errors as in Figure \ref{fig:SAbasic}}
	\label{fig:SAbasicmass}
\end{figure*}

Figure \ref{fig:SAbasicmass} shows the relationship between $r/R_{\rm vir}$ and $\cos\theta$, subsampled by subhalo to host halo mass ratio, $M_{\rm sub}/M_{\rm host}$. The upper panels show the 2D histograms and the lower panels show the excess probability density of the subsample.
As indicated by the median values of $\cos\theta$ at each $r/R_{\rm vir}$ and the excess probability density, the spin alignment between subhalos and host halos becomes more pronounced as the mass ratio increases.

The radial distribution of subhalos depends strongly on the mass ratio. As indicated by the color scale in the 2D histogram, subhalos with small mass ratios $10^{-3}\leq M_{\rm sub}/M_{\rm host}<10^{-2}$ are primarily located in the outskirts of the host halo. In contrast, those with large mass ratios $10^{-1}\leq M_{\rm sub}/M_{\rm host}<1$ are concentrated around $r/R_{\rm vir} \sim 0.4$ and $\cos\theta \sim 1$.
This is because subhalos with higher mass ratios relative to their host halos experience stronger dynamical friction, causing them to sink toward the center on shorter timescales.
Consequently, the alignment observed in the central regions in Figure \ref{fig:SAbasic} (without mass-ratio binning) is dominated by the contribution of high-mass-ratio subhalos.

\subsection{Subhalo Spin Alignment and Orbital Motion}\label{subsec:orbital}
\begin{figure*}[ht!]
	\centering
	\includegraphics[width=2\columnwidth]{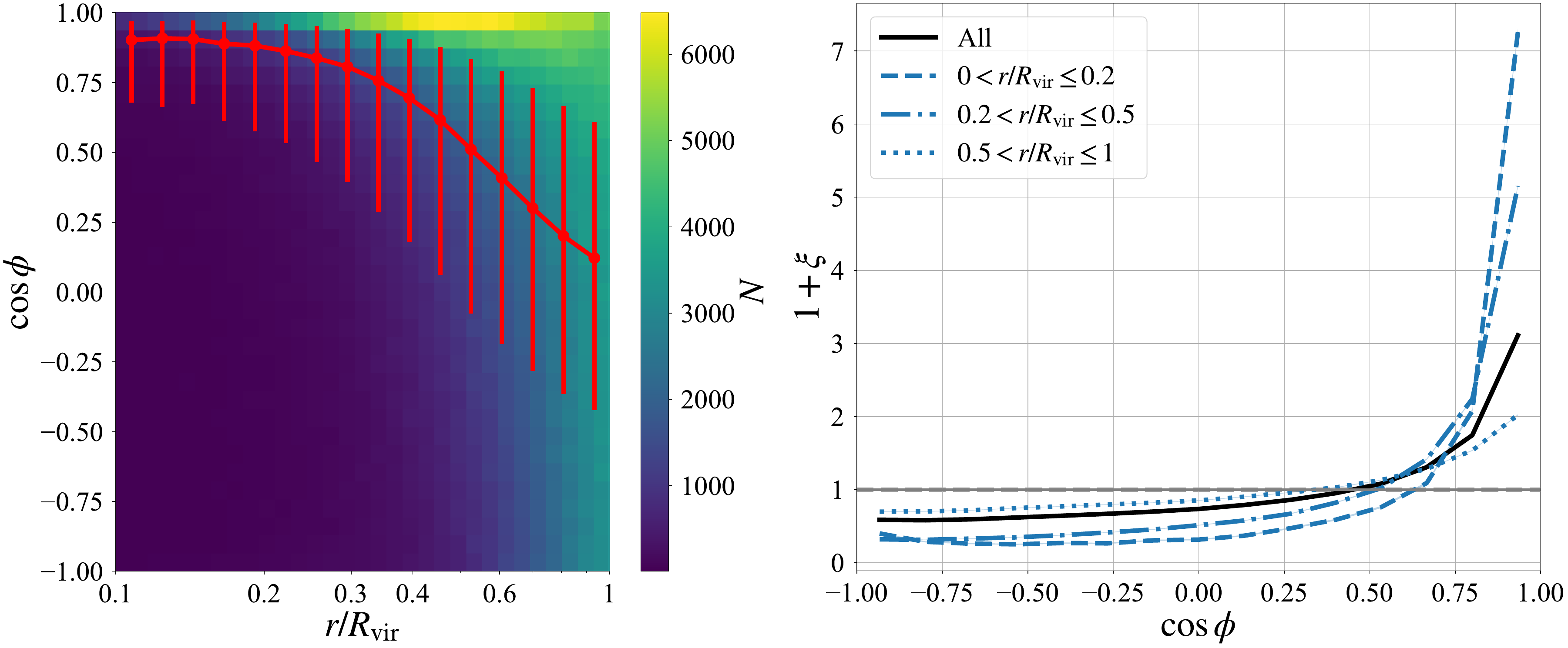}
	\caption{Same as Figure \ref{fig:SAbasic}, but for the alignment between subhalo spin and orbital angular momentum, $\cos\phi$.}
	\label{fig:SAjsubjorb}
\end{figure*}

In the previous section, a tendency for the spin of subhalos and host halos to align was observed. If this alignment originates from tidal torques exerted by the host halo \citep{2004MNRAS.352..376A,2007MNRAS.375..489H,2011MNRAS.413.3013L}, the spin of the subhalos is expected to align with the direction of their orbital angular momentum. We thus investigate the alignment between the spin and orbital angular momentum of the subhalo, $\cos\phi$. Figure \ref{fig:SAjsubjorb} is formatted similarly to Figure \ref{fig:SAbasic}, but the vertical axis of the 2D histogram and the horizontal axis of the excess probability density represent the angle between spin and orbital angular momentum of subhalo, $\cos\phi$.
Figure \ref{fig:SAjsubjorb} shows that, similarly to Figure \ref{fig:SAbasic}, the alignment signal increase toward the center of the host halo.
Furthermore, the excess probability density values for $\cos\phi > 0.5$ exceed those for $\cos\theta > 0.5$, indicating that the alignment between subhalo orbital angular momentum and spin is stronger than the alignment between subhalo and host halo spins. The alignment between subhalo spin and orbital angular momentum is physically consistent with a scenario where subhalos acquire their spin via tidal torquing within the host halo environment.

\begin{figure*}[ht!]
	\centering
	\includegraphics[width=2\columnwidth]{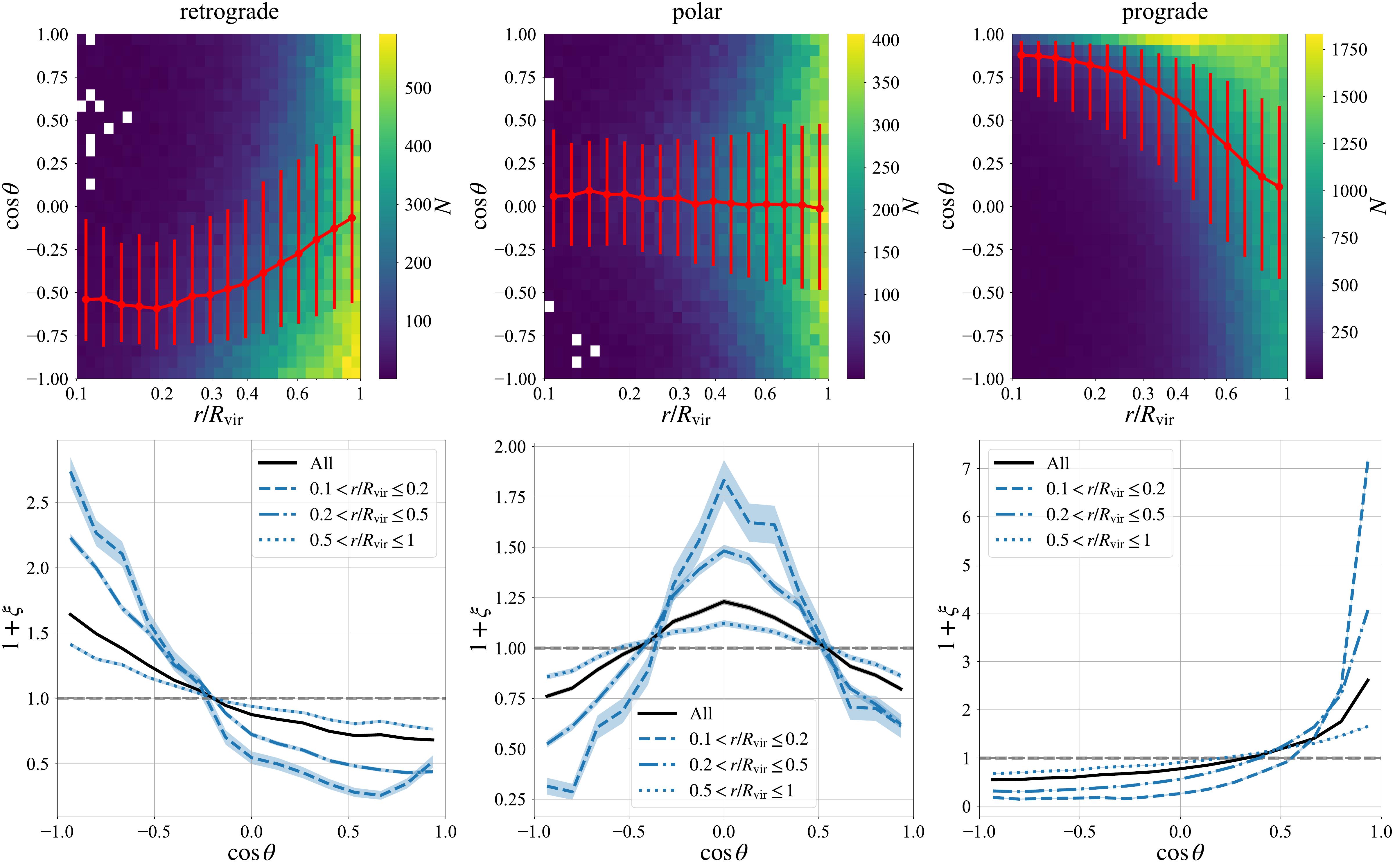}
	\caption{Upper panels: 2D histograms of $\cos\theta$ versus $r/R_{\rm vir}$. Lower panels: Corresponding excess probability density of $\cos\theta$. The sample is restricted to subhalos located near the equatorial plane of the host halo. The columns show results for different orbital orientation: retrograde (left, $-1 \le \cos\eta \le -1/3$), polar (middle, $-1/3 \le \cos\eta < 1/3$) and prograde (right, $1/3 \le \cos\eta < 1$). Here, $\cos\eta$ is the alignment between the subhalo orbital angular momentum $\vec{J}_{\rm orb}$ and the host halo spin $\vec{J}_{\rm host}$.}
	\label{fig:SAorbital}
\end{figure*}

If this scenario is correct, the trend of spin alignment is expected to vary depending on the orbital orientation of the subhalos. We therefore investigate the relationship between $\cos\theta$ and $r/R_{\rm vir}$ by categorizing subhalos into different orbital types.
We classify subhalos located near the equatorial plane of the host halo ($|\cos\chi| < 0.5$) into three orbital categories, based on the alignment between their orbital angular momentum and that of the host halo's spin ($\cos\eta$).
These categories are defined as retrograde ($-1 \leq \cos\eta < -1/3$), polar ($-1/3 \leq \cos\eta \leq 1/3$), and prograde ($1/3 < \cos\eta \leq 1$).
These orbital subsamples account for $11.9\%$ (retrograde), $8.56\%$(polar) and $36.9\%$(prograde) of the full sample.
A classification based solely on $\cos\eta$ cannot distinguish whether a subhalo at $\cos\chi \approx 1$ is in a genuine polar orbit or is a subhalo from a different orbit that is only transiently passing through the polar region.
By restricting the sample to the region near the equatorial plane ($|\cos\chi| < 0.5$), we enable a robust classification of these orbits. 
Additionally, subhalos near the host center undergo merging processes with the host core and their position vectors $\vec{X}_{\rm sub}$ approach zero, making it difficult to accurately calculate $\vec{J}_{\rm orb}$. We therefore impose the constraint $r/R_{\rm vir} > 0.1$ to ensure the reliability of our orbital calculations.
Figure \ref{fig:SAorbital} presents 2D histograms of $\cos\theta$ and $r/R_{\rm vir}$ and the excess probability density of $\cos\theta$ for three orbital types, retrograde, polar, and prograde.

The 2D histogram for subhalos on prograde orbits displays a clear tendency for $\cos\theta$ to increase toward the host halo center, mirroring the global trend seen in Figure \ref{fig:SAbasic}. As shown in the excess probability density of prograde sample, this alignment is even more pronounced than that of the entire subhalo sample. In contrast, for polar and retrograde orbits, $\cos\theta$ values tend toward 0 and $-1$, respectively, as they approach the center.
This provides strong evidence for the tight alignment between spin and orbital angular momenta, as shown in Figure \ref{fig:SAjsubjorb}, indicating that subhalos acquire their spin through tidal torques from the host halo.
Since host halos gain angular momentum through matter accretion, prograde orbits are statistically more common than retrograde ones \citep{2014MNRAS.443.1274L, 2025MNRAS.538..963M}, leading to the emergence of the identified global spin alignment trend.
In contrast, subhalos on polar and retrograde orbits acquire spins that are perpendicular and opposite to the host halo's spin direction, respectively.
It is known that most subhalos with large mass ratios are prograde orbits \citep{2021ApJ...914...86A,2025MNRAS.538..963M}, and we verified similar results. In Figure \ref{fig:SAbasicmass}, the trend was more pronounced for subhalos with larger mass ratios, suggesting that this behavior reflects the influence of their orbital angular momentum.

\subsection{Dependence on accretion redshift of subhalos}

Based on the tidal torquing scenario, subhalos accreted at earlier epochs are expected to exhibit more pronounced spin alignment due to prolonged tidal interactions with the host halo. 

\begin{figure*}[ht!]
	\centering
	\includegraphics[width=2\columnwidth]{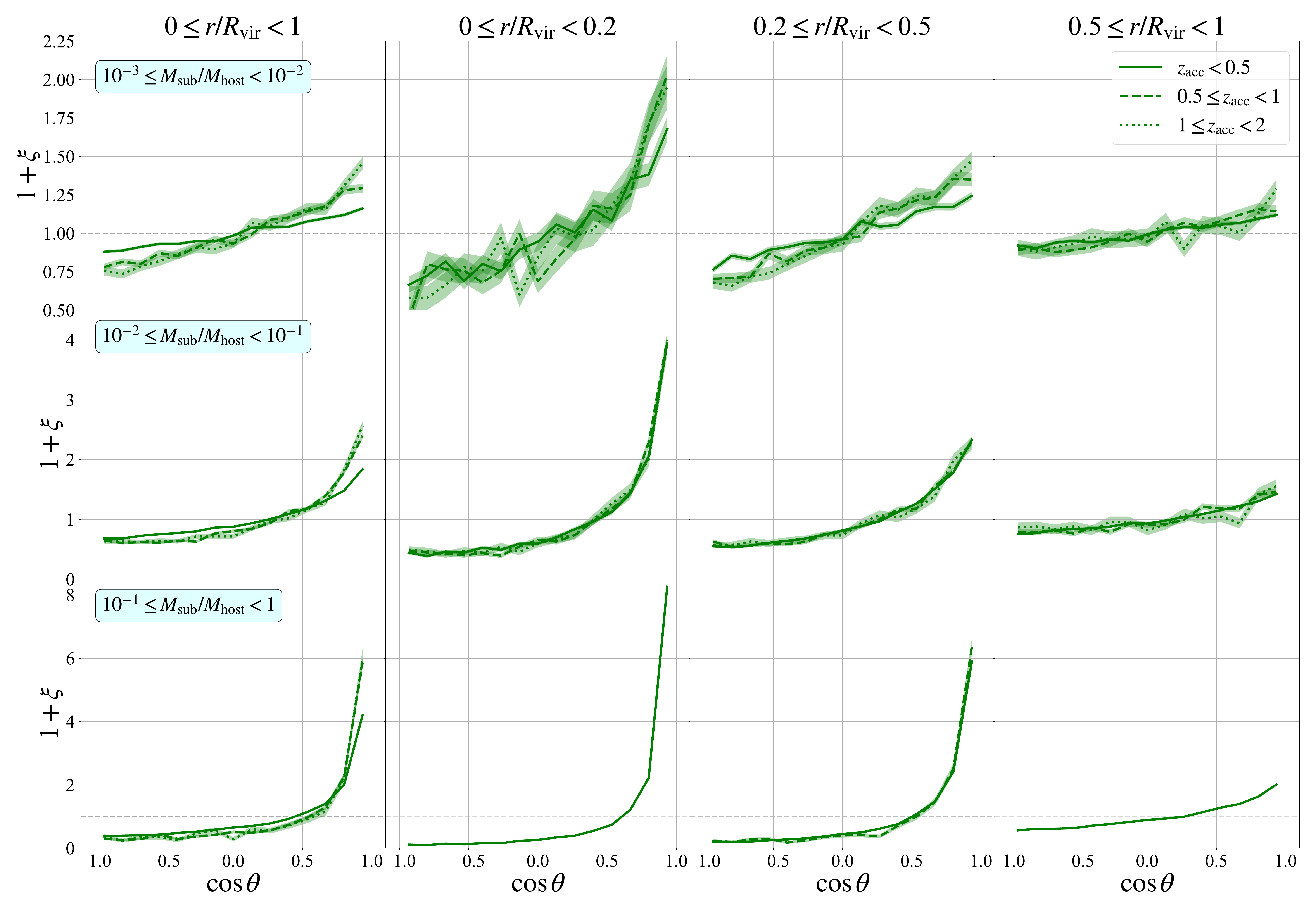}
	\caption{Excess probability density of $\cos\theta$ for subhalos. In each panel, the lines represent three categories of the accretion redshift $z_{\rm acc}$ with more than 1,000 subhalos, with Poisson errors. The panels are arranged in rows by mass ratio $M_{\rm sub}/M_{\rm host}$ and in columns by radial distance from the host halo center ($r/R_{\rm vir}$). Rows, from top to bottom, correspond to $10^{-3} \le M_{\rm sub}/M_{\rm host} < 10^{-2}$, $10^{-2} \le M_{\rm sub}/M_{\rm host} < 10^{-1}$, and $10^{-1} \le M_{\rm sub}/M_{\rm host} < 1$. Columns, from left to right, correspond to the full radial range ($0 \le r/R_{\rm vir} < 1$) and three radial subsamples: $0 \le r/R_{\rm vir} < 0.2$, $0.2 \le r/R_{\rm vir} < 0.5$, and $0.5 \le r/R_{\rm vir} < 1$. For $10^{-1} \le M_{\rm sub}/M_{\rm host} < 1$, all $z_{\rm acc}$ bins in the full radial range sample contain more than 1,000 subhalos. However, when the sample is further divided into radial subsamples, subsamples with more than 1,000 subhalos are found only for $0 \le z_{\rm acc} < 0.5$, and for $0.5 \le z_{\rm acc} < 1$ in the radial bin $0.2 \le r/R_{\rm vir} < 0.5$.}
	\label{fig:HistSAjsubjhost_zacc_distance}
\end{figure*}

Figure \ref{fig:HistSampleZacc} shows the differential histogram of subhalos as a function of the accretion redshift, $z_{\rm acc}$. The number of subhalos decreases toward higher $z_{\rm acc}$, particularly in higher mass-ratio subsamples, leading to limited sample sizes at high $z_{\rm acc}$, especially when further subdivided by mass ratio and radial distance.
Figure \ref{fig:HistSAjsubjhost_zacc_distance} shows the excess probability density of $\cos\theta$ for subhalos as a function of $z_{\rm acc}$. Results are presented for subsamples divided by $M_{\rm sub}/M_{\rm host}$ and $r/R_{\rm vir}$ to isolate the effect of $z_{\rm acc}$ by removing their dependence on these properties. Only subsamples with more than 1,000 subhalos are included to ensure statistical robustness.
For the highest mass-ratio range ($10^{-1} < M_{\rm sub}/M_{\rm host} < 1$), most radial subsamples in $0.5 < z_{\rm acc} < 1$ and $1 < z_{\rm acc} < 2$ contain fewer than 1,000 subhalos.

The full radial range sample ($0 \leq r/R_{\rm vir} < 1$) shows a dependence on $z_{\rm acc}$ across all mass-ratio subsamples. In particular, the low-$z_{\rm acc}$ subsample ($0 \leq z_{\rm acc} < 0.5$) exhibits weaker alignment than the higher-$z_{\rm acc}$ subsamples.
For low-mass-ratio subhalos ($10^{-3} \leq M_{\mathrm{sub}}/M_{\mathrm{host}} < 10^{-2}$), the alignment is most pronounced in the high-$z_{\rm acc}$ subsample ($1 \leq z_{\rm acc} < 2$), while the lower-$z_{\rm acc}$ subsamples exhibit weaker alignment.
For intermediate- and high-mass-ratio subhalos ($10^{-2} \leq M_{\mathrm{sub}}/M_{\mathrm{host}} < 1$), the low-$z_{\rm acc}$ subsample shows the weakest alignment, while the $0.5 \leq z_{\rm acc} < 1$ and $1 \leq z_{\rm acc} < 2$ subsamples exhibit similar levels of alignment.

However, the dependence on $z_{\rm acc}$ is significantly reduced for $10^{-2} < M_{\rm sub}/M_{\rm host} < 10^{-1}$ subsamples when the analysis is performed within individual radial bins. This suggests that the $z_{\rm acc}$ dependence observed across the full radial range may primarily be an apparent effect arising from the radial distribution of subhalos. 
For the highest-mass-ratio range ($10^{-1} < M_{\rm sub}/M_{\rm host} < 1$), the number of comparable samples is limited, but in the radial bin $0.2 \leq r/R_{\rm vir} < 0.5$, the distributions for $0.5 \leq z_{\rm acc} < 1$ and $1 \leq z_{\rm acc} < 2$ are nearly indistinguishable.
This indicates that the apparent dependence arises from the mixing of subhalos at different radial positions.
In contrast, for low mass-ratio subhalos, little to no dependence is observed in the outer regions ($0.5 < r/R < 1$), whereas a clear dependence remains in the inner regions ($0 < r/R < 0.2$ and $0.2 < r/R < 0.5$), indicating that the alignment depends not only on radial position but also on $z_{\rm acc}$.

The dependence on $z_{\rm acc}$ seen in intermediate- and high-mass-ratio subhalos in the full radial range is consistent with the effects of dynamical friction. The timescale for a subhalo to sink toward the center of the host halo is governed by dynamical friction and becomes shorter for subhalos with larger mass ratios. Subhalos accreted at higher $z_{\rm acc}$ have had sufficient time to migrate toward the inner regions of the host halo. Since the alignment signal is intrinsically stronger in the central regions, this leads to an apparent enhancement of the alignment when integrated over the full radial range.
In contrast, for low-mass-ratio subhalos, the dependence on $z_{\rm acc}$ does not disappear even within individual radial bins, suggesting that the accretion history itself may exert an intrinsic influence on subhalo alignment, while the physical origin of this dependence remains unclear. Furthermore, $z_{\rm acc}$ and $r/R_{\rm vir}$ are inherently correlated, making it challenging to disentangle their individual contributions.

\section{DISCUSSION AND CONCLUSION}

In this study, we used the Shin-Uchuu cosmological $N$-body simulations to investigate the relationship between the distance from the center of the host halo and the spin direction of subhalos, focusing on subhalos surviving at $z=0$ and tracing their accretion history back to $z=2$.
Statistically, the spin direction of subhalos exhibits an increasing tendency to align with that of their host halo as the distance to the host center decreases.
This trend is driven by the fact that the orbital angular momentum of most subhalos aligns with the host halo's spin direction. As subhalos acquire spin in the same direction as their orbital angular momentum via tidal field, they consequently tend to align with the spin of the host halo. Therefore, if the orbital motion of a subhalo is misaligned with the rotation of the host halo, such as in retrograde or polar orbits, the subhalo's spin does not align with that of the host halo.

These findings provide robust evidence that subhalos acquire their spin through tidal torques from their host halos.
Although a dependence on the accretion redshift ($z_{\rm acc}$) is statistically detected in the full radial range sample, our analysis indicates that this trend is closely linked to the radial distribution of subhalos. Subhalos accreted earlier have had more time to migrate inward and are therefore statistically expected to reside closer to the host halo center, naturally giving rise to a correlation between $z_{\rm acc}$ and $r/R_{\rm vir}$. Since these quantities are closely linked, it is difficult to disentangle the effects of accretion history from those of the current radial position.

For high mass-ratio subhalos, a dependence on $z_{\rm acc}$ is observed when considering the full radial range. However, this dependence largely disappears when the analysis is performed within individual radial bins. This indicates that the apparent $z_{\rm acc}$ dependence is primarily driven by the radial distribution of subhalos.
In contrast, for low mass-ratio subhalos, a residual dependence on $z_{\rm acc}$ remains in the inner regions. This suggests that not only the radial distance from the host halo center but also $z_{\rm acc}$ may contribute to the acquisition of subhalo spin. These results highlight the importance of further investigating the timescales of spin acquisition through tidal torquing.

While subhalos accreted earlier are statistically expected to be located near the host center, suggesting an inherent link between $z_{\rm acc}$ and $r/R_{\rm vir}$, caution is required when interpreting the $z_{\rm acc}$ dependence across the entire population. This is because subhalos that migrated into the innermost regions are susceptible to being disrupted by the host halo. Such disruption leads to a potential survival bias in the remaining population, which may artificially mask or alter the underlying physical correlations between accretion history and alignment.

While subhalo major axes tend to point toward the host halo center as they approach it, a misalignment is often found in the innermost regions ($r/R_{\rm vir} \lesssim 0.2$) \citep{2008ApJ...672..825P,2008MNRAS.388L..34K,2020MNRAS.495.3002K}. Our findings may help elucidate the physical mechanisms behind this central misalignment, thereby offering more profound insights into the intrinsic alignment of halos and galaxies.

If the misalignment of the subhalo major axes is indeed linked to subhalo spin, uncovering the underlying physical mechanisms will contribute to refining the probability distributions of major-axis alignments, thereby enhancing the precision of alignment models that incorporate subhalo alignment \citep{2024OJAp....7E..45V}.

Additionally, galaxies with low surface brightness, called ultra-diffuse galaxies (UDGs), have been observed to exhibit a tendency for high spin parameters \citep{2023MNRAS.522.1033B,2017MNRAS.470.4231R,2016MNRAS.459L..51A}. Investigating the influence of tidal torques on galaxy evolution may therefore help constrain the formation scenarios of UDGs.

However, since massive satellites like the LMC significantly influence subhalo orbits \citep{2023Galax..11...59V} and anisotropic accretion from cosmic filaments dictates the orientation of host halo spins \citep{2012MNRAS.421L.137L,2020MNRAS.495..502M,2018MNRAS.481..414G,2021MNRAS.503.2280G}, a detailed investigation of spin alignment requires a multiscale approach accounting for both cosmic filaments and individual subhalo interactions. Additionally, the spin of a galaxy does not necessarily coincide with that of its host dark matter halo \citep{2010MNRAS.404.1137B, 2023MNRAS.518.1002R}. Therefore, comparing the spin acquisition of baryonic and dark matter components via hydrodynamic simulations is an essential next step.

This study has provided evidence that subhalos acquire spin through tidal torque exerted by their host halos, aligning their rotation with their orbital angular momentum. Statistically, it shows a tendency for the spin directions of subhalos and host halos to align.
These findings advance our understanding of satellite galaxy formation and, by characterizing the nature of intrinsic alignment, facilitate the refinement of weak lensing analysis while mitigating associated systematic uncertainties.

\begin{acknowledgments}
This work was supported by JSPS/MEXT KAKENHI Grant Numbers JP25H00671 (KW) and JP25H00662 (TI).
This work was also supported by JST SPRING Grant Number JPMJSP2119,by MEXT as a “Program for Promoting Researches on the Supercomputer Fugaku” (Toward a unified view of the Universe: from large-scale structures to planets, Grant Number JPMXP1020200109) and Program for Promoting Researches on the Supercomputer Fugaku” (Structure and Evolution of the Universe Unraveled by Fusion of Simulation and AI; Grant Number JPMXP1020230406). TI has been supported by IAAR Research Support Program in Chiba University Japan, and JICFuS.
We thank Instituto de Astrofisica de Andalucia (IAA-CSIC), Centro de Supercomputacion de Galicia (CESGA) and the Spanish academic and research network (RedIRIS) in Spain for hosting Uchuu DR1 and DR2 in the Skies \& Universes site for cosmological simulations. The Uchuu simulations were carried out on Aterui II supercomputer at Center for Computational Astrophysics, CfCA, of National Astronomical Observatory of Japan, and the K computer at the RIKEN Advanced Institute for Computational Science. The Uchuu DR1 and DR2 effort has made use of the skun@IAA\_RedIRIS and skun6@IAA computer facilities managed by the IAA-CSIC in Spain (MICINNEU-Feder grant EQC2018-004366-P).
\end{acknowledgments}

\bibliography{export-bibtex}{}
\bibliographystyle{aasjournalv7}

\end{document}